\documentclass[journal]{IEEEtran}
\usepackage{graphicx}
\usepackage{epstopdf}
\usepackage{cite}
\usepackage{amsfonts}
\usepackage{amssymb}
\usepackage{amsmath}
\usepackage{mathrsfs}
\usepackage{bm}
\usepackage{amsmath}
\usepackage{mdwmath}
\usepackage{mdwtab}
\usepackage[ruled,vlined,linesnumbered]{algorithm2e} 
\usepackage{subfigure}

\usepackage{makecell,multirow,diagbox}
\newcommand{\tabincell}[2]{\begin{tabular}{@{}#1@{}}#2\end{tabular}}

\begin{document}

\title{Exploiting Wireless Channel State Information Structures Beyond Linear Correlations: A Deep Learning Approach}
\author{Zhiyuan Jiang, Sheng Chen, Andreas F. Molisch, \IEEEmembership{Fellow IEEE}, Rath Vannithamby, Sheng Zhou, \\and Zhisheng Niu, \IEEEmembership{Fellow IEEE}
\thanks{Zhiyuan Jiang is with %Key laboratory of Specialty Fiber Optics and Optical Access Networks, Joint International Research Laboratory of Specialty Fiber Optics and Advanced Communication, %
	Shanghai Institute for Advanced Communication and Data Science, Shanghai University, Shanghai 200444, China. 
	Sheng Chen, Sheng Zhou and Zhisheng Niu are with Beijing National Research Center for Information Science and Technology, Tsinghua University, Beijing 100084, China. 
	Andreas F. Molisch is with the Ming Hsieh Department of Electrical Engineering, University of Southern California, Los Angeles, CA 90089, USA. 
	Rath Vannithamby is with the Intel Corporation, Hillsboro, OR 97124, USA.
	
	This work is sponsored in part by the Nature Science Foundation of China (No. 61701275, No. 91638204, No. 61861136003, No. 61571265, No. 61621091), National Key R\&D Program of China 2018YFB0105005, and Intel Collaborative Research Institute for Mobile Networking and Computing. The corresponding author is Sheng Zhou.
}}
\maketitle

\begin{abstract}
Knowledge of information about the propagation channel in which a wireless system operates enables better, more efficient approaches for signal transmissions. Therefore, channel state information (CSI) plays a pivotal role in the system performance. The importance of CSI is in fact growing in the upcoming 5G and beyond systems, e.g., for the implementation of massive multiple-input multiple-output (MIMO). However, the acquisition of timely and accurate CSI has long been considered as a major issue, and becomes increasingly challenging due to the need for obtaining CSI of many antenna elements in massive MIMO systems. To cope with this challenge, existing works mainly focus on exploiting linear structures of CSI, such as CSI correlations in the spatial domain, to achieve dimensionality reduction. In this article, we first systematically review the state-of-the-art on CSI structure exploitation; then extend to seek for deeper structures that enable remote CSI inference wherein a data-driven deep neural network (DNN) approach is necessary due to model inadequacy. We develop specific DNN designs suitable for CSI data. Case studies are provided to demonstrate great potential in this direction for future performance enhancement.
\end{abstract}

\section{Introduction}
\label{sec_intro}
Channel state information (CSI) is a characterization of the wireless propagation channels between transmitters and receivers, whose importance has been growing significantly for wireless systems. The early radio systems which employ non-coherent demodulation schemes, e.g., amplitude or frequency modulations, do not really need CSI. However, with the demand for high data-rate services and the scarcity of wireless spectrum, more advanced transmission schemes are adopted for which CSI is necessary. Specifically, modulation schemes such as quadrature amplitude modulation (QAM) that require CSI for coherent demodulation become dominant; The wideband and spatial signal transmission, which are the major advancements of 2G-3G and 4G-5G cellular systems respectively, both require an increasing amount of CSI. In particular, CSI is indispensable for the upcoming 5G and beyond systems which heavily rely on the spatial signal processing of massive multiple-input multiple-output (M-MIMO)---a main enabler of several key performance targets for 5G. CSI-assisted spatial multiplexing brings enormous extra degree-of-freedoms (DoFs) for high data-rate applications, whereby the same time and frequency resource elements are reused such that the channel capacity may increase by orders of magnitude. The transmission of millimeter-wave signals relies on leveraging the high beamforming gain, which again requires accurate CSI. Moreover, for ultra-reliable and low-latency communication (URLLC), CSI offers robustness against multi-path delay dispersion and errors due to channel fluctuations. The requirement for CSI goes beyond spatial signal processing, and is mandatory for user scheduling, handover planning, buffer design for wireless video and etc.
\begin{table*}[!t]
	\centering
	\caption{Summary of State-of-the-art on CSI structure exploitation}
	\label{table_review}
	\begin{tabular}{|c|c|c|c|c|c|}
		\hline
		Ref. & Domain & Prior knowledge & Correlation model & Technique & Rationality \\ 
		\hline
		\cite{Adhikary13} & \multirow{6}*{Spatial domain} & \multirow{2}*{CCMs of users} & \multirow{2}*{Karhunen-Lo\`{e}ve model} & Two-layer beamforming & \multirow{6}*{\tabincell{c}{\textrm{Sparse angular}\\\textrm{power spectrum}}}  \\
		\cline{1} \cline{5}
		\cite{Jiang14} &  &  &  & \tabincell{c}{\textrm{Optimized pilots and}\\\textrm{feedback codebook}} &   \\
		\cline{1} \cline{3-5}
		\cite{brady13} &  & ---- & \multirow{3}*{Angular representation} & Lens-antenna-array &   \\
		\cline{1} \cline{3} \cline{5}
		\cite{molisch04_as} &  & ---- &  & Butler matrix &   \\
		\cline{1} \cline{3} \cline{5}
		\cite{Rao14} &  & Common scatterers &  & Compressive sensing &   \\
		\hline
		\cite{4} & \multirow{3}*{Time/Doppler domain} & Channel coherence time & Block fading & Lyapunov optimization & \multirow{3}*{\tabincell{c}{\textrm{Channel coherence}\\\textrm{over time}} }  \\
		\cline{1} \cline{3-5}
		\cite{had17} & & ---- & Doppler spectrum  & OTFS &   \\
		\cline{1} \cline{3-5}
		\cite{Choi142} &  & Time correlation coefficients & Gauss-Markov model & Karman filter &   \\
		\hline
		\cite{gao16_cs} & \multirow{3}*{Frequency domain} & Common sparsity assumption & OFDM-based & Compressive sensing & \multirow{3}*{Sparse delay profile}  \\
		\cline{1} \cline{3-5}
		
		\cite{you16} &  & Angle-delay profile & OFDM-based & Phase-shifted pilots &   \\
		\cline{1} \cline{3-5}
		\cite{ven17} &  & Common sparsity assumption & OFDM and SC-FDE & Compressive sensing & \\
		
		\hline
	\end{tabular}
\end{table*}

However, the acquisition of timely and accurate CSI entails considerable air-interface resource overhead. It is fair to say that the CSI acquisition (one-shot) overhead scales, at least, with the channel DoF, which is the smaller of the numbers of transmit antennas and sum of user antennas, no matter time-division-duplex (TDD) or frequency-division-duplex (FDD) system is considered. With high channel DoF requirement, the overhead constitutes a major hurdle to realize the promised performance. Currently, pilot overhead consumes about $19$\% of system resources in the current $8$-antenna LTE systems and the percentage will grow significantly in the future.

The CSI overhead issue has attracted broad attention in the literature. It was discovered that exploiting linear correlations of CSI among, e.g., co-located antennas, different time instances and different frequency subcarriers, can alleviate this problem. In this article, we systematically review these existing works, and then point out that there are unexplored frontiers in the CSI structure that may bring tremendous additional performance benefit in scenarios conventionally believed not to be amenable to such improvements. Case studies based on applying deep neural networks (DNNs) are also conducted to validate the feasibility and potentials. It is argued that, unlike other modules in wireless communications which usually have well-accepted models, deep learning---a powerful tool \cite{lecun15} that utilizes DNN with huge recent success for image and speech data---is necessary for the \emph{inadequate-modeled} CSI structure exploitation problem.

\section{Exploiting CSI Structure: State-of-the-Art}
\label{sec_review}
The reason for investigating the structures of CSI is to achieve \emph{dimensionality reduction}. In its most general form, CSI describes the time-varying channel impulse responses of each transmit-receive antenna pair within the system bandwidth. Assuming an OFDM-based system, the CSI can be defined as an $M \times N \times T \times B$ dimensional tensor where $M$ and $N$ are the numbers of base station (BS) antennas and sum of user antennas, respectively, and the numbers of OFDM symbols and subcarriers are denoted by $T$ and $B$, respectively. Each dimensionality of CSI requires at least one estimate (consequently overhead without considering shared pilots received at multiple antennas) if no underlying structures of CSI are exploited. Fortunately, it has been shown that dimensionality reduction is possible in practical scenarios by leveraging the CSI structures, most simply linear correlations. The state-of-the-art on CSI structure exploitation is summarized in Table \ref{table_review}. In essence, the key idea of existing methods is to leverage the signal sparsity in the transform domains, such as delay, Doppler and angular domains, and then apply Nyquist sampling (Doppler domain) or exploit linear correlation structure in those domains. Specific examples are illustrated as follows.

\emph{\textbf{Spatial domain}}: The majority of the works focus on the spatial domain CSI linear correlations exploitation, since it has been discovered by field measurements that the angle-of-arrival (AoA) spectrum of multi-path components (MPCs), which is the support of linear basis for spatial CSI, is sparse compared to the angular range covered by the BS antenna array. Therefore, by transforming the CSI into the angular domain where most of the angular bins carry negligible energy, the dimensionality of CSI is significantly reduced. To gain knowledge on exact sparsity patterns in the angular domain, there is work \cite{Adhikary13, Jiang14} assuming prior knowledge of the channel correlation matrices (CCMs). Adhikary \emph{et al.} \cite{Adhikary13} propose a two-layer beamforming approach whereby the CCMs are first exploited for spatial division and then the instantaneous CSI is used for fine-grained spatial multiplexing. It is shown that by optimizing the training and feedback procedures considering the spatial correlations, the achievable rates of FDD systems even approach TDD systems with tens of BS antennas \cite{Jiang14}. Leveraging a lens-antenna-array \cite{brady13}, which transforms the signals into angular domain by a lens for wireless signals, or equivalently a Butler matrix \cite{molisch04_as}, the dimensionality reduction is achieved with little extra signal processing. Without knowledge of the CCMs, the spatial sparsity pattern is unknown; compressive sensing (CS) based approaches are consequently proposed \cite{Rao14} to address this issue. 

\emph{\textbf{Time, Doppler and frequency domains}}: The simplest and most widely-used time correlation model is the block fading model, based on which the CSI dimensionality can be reduced by a factor of the block length. The heterogeneous block lengths case is investigated in \cite{4} where the pilots for different users are scheduled according to their traffic demands and block lengths. The time correlation can also be exploited in an FDD system by either letting users combine the current channel outputs with historic ones for channel estimations, or feeding back the optimal pilot sequences \cite{Choi142}. Ref. \cite{had17} proposes an efficient pilot packing scheme which is enabled by orthogonal time frequency space (OTFS) modulation. It arranges pilots in the delay-Doppler domain and thus can inherently take into account the different timescales (support of Doppler dispersion) of different users. Frequency domain dimensionality reduction is enabled by sparse delay profile which has been extensively investigated in, e.g., OFDM systems. Recent efforts focus on utilizing the joint angular-delay sparsity based on the CS framework. The sparse signal support in the delay domain is estimated by noticing that the BS antennas share the common sparse pattern \cite{gao16_cs}. Training sequence optimization with adjustable phase shift pilots is investigated in \cite{you16}. The pilots are phase shifted such that they are non-overlapped in the delay-spatial space and hence the overall pilot resource occupation is reduced. Joint frequency and spatial correlations are treated in the context of channel estimations in \cite{ven17}.

\section{Beyond Linear Structures: Theoretical Study and DNN-Based Implementation}
\label{sec_new}
Whereas different domains have been considered by existing works, most have been focused on linear correlations such as sparse spatial steering vectors or frequency response, and Gauss-Markov time correlations. From a data statistics perspective, CSI in its general form can potentially have non-linear structures waiting to be explored. In this section, we first present a theoretical study of the non-linear CSI structures based on mutual information (MI) and Cram\'er-Rao lower bound (CRLB), indicating the existence of such structures and theoretical limits (for a simplified model) in terms of CSI inference accuracy. Then a DNN-based remote CSI inference framework is introduced to exploit these structures.
\subsection{Understanding Non-Linear CSI Structures}
\label{sec_linear}
It is well known that in calculus, the value increment of a smooth function caused by the increment of the arguments can be well approximated by a linear function of the argument increment when the increment is small. However, the approximation becomes loose with the argument increment growing larger and that is when non-linear representations are needed. A similar principle applies to the CSI. In each domain, the linear correlations are only effective within their corresponding scales. For instance, it is widely accepted that only CSIs of antennas whose separations are on the order of wavelengths are linearly correlated. The \emph{region of stationarity}, which could be in spatial, time or frequency domain, is hence defined, within which the CSI can be treated as a wide-sense stationary process. Beyond the region of stationarity, such simple linear structures no longer exist. For example, CSIs of antennas at geographically separated sites are considered linearly independent. Similarly, CSIs of paired frequency bands in FDD systems, or CSIs separated by long time intervals, are also considered linearly independent. 
\begin{figure}[!t]
    \centering  
    {\includegraphics[width=0.5\textwidth]{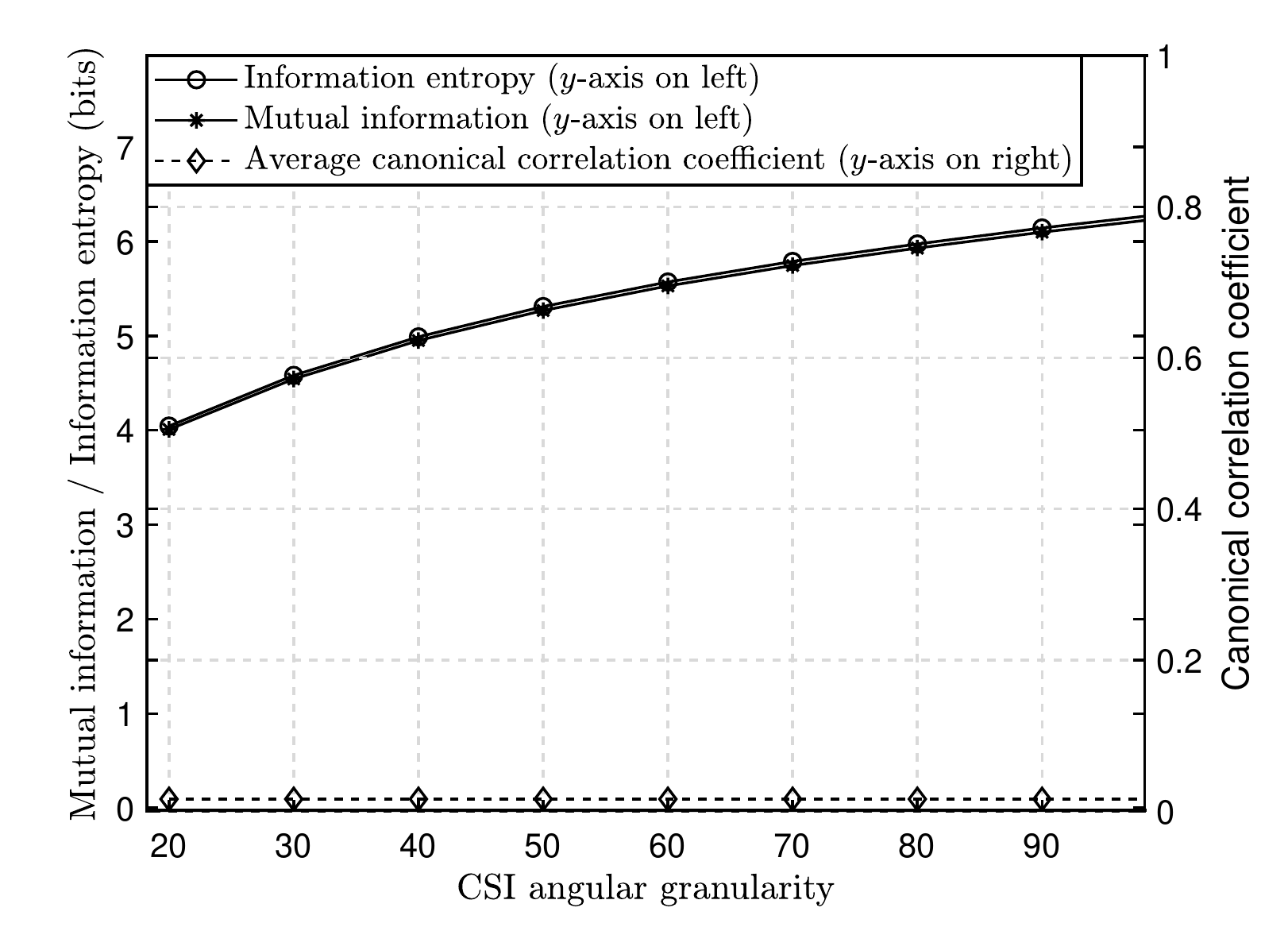}}
    \caption{Mutual information analysis on the relationship between local CSI and remote CSI based on the Wireless Insite ray-tracing propagation software. The average canonical correlation coefficient characterizes the linear correlation.}  
    \label{sim_info} 
\end{figure} 
    
Nevertheless, even though the CSIs beyond the regions of stationarity may not be linearly correlated, they can be related to each other through more complex connections. Before diving into the exploitation of such structures, in this section, the feasibility issue is first investigated. An imminent question is whether there exist such richer structures of CSI. In theory, the linear correlations therein, termed as Pearson correlations in statistics, can express the CSI structures completely only when the CSI is multivariate Gaussian distributed as in the rich scattering environment. Generally speaking, the information given by linear correlations is insufficient to define the dependency structure between random variables. Moreover, the rich scattering condition is seldom perfectly satisfied given realistic limited-scattering propagation environment of, for example, millimeter-wave systems with dominant line-of-sight (LoS) MPCs. The non-linear relevance of CSIs at two remote sites outside their respective region of stationarity can emerge due to common scatterers and location-based LoS components. To validate this conclusion with more concrete evidence, we adopt the mutual information (MI) as a metric to measure the dependency between CSIs since the MI is capable of detecting almost any functional dependency among random variables based on sampled data. The MI between the CSI vectors of a single-antenna user at two remote BS sites is calculated by collecting the CSI statistics generated by a Wireless Insite ray-tracing software. In Fig. \ref{sim_info}, we consider a scenario where there are one macro BS (MBS) with $100$ antennas forming a uniform linear array (ULA) and one small BS (SBS) with $20$ ULA antennas. The CSIs at two BSs are both quantized to avoid complications about MI calculation of continuous random variables. It is shown that the MI between two remote CSI vectors is very close to the information entropy of the CSI at the SBS under the discrete Fourier transform (DFT) quantization codebook, meaning that the CSI at the SBS is strongly dependent on the CSI at the MBS. On the other hand, the linear correlation between these two CSIs is approximately zero. The linear correlation shown in the figure is measured by the average canonical correlation coefficient which is used for linear correlation analysis on random vectors. This finding confirms the fact that rich, non-linear structures exist in the CSI, and more importantly, materializes the potential for remote CSI inference by which the CSI can be inferred by the CSI at a remote site.
\begin{figure*}[!t]
	\centering
	\includegraphics[width=0.8\textwidth]{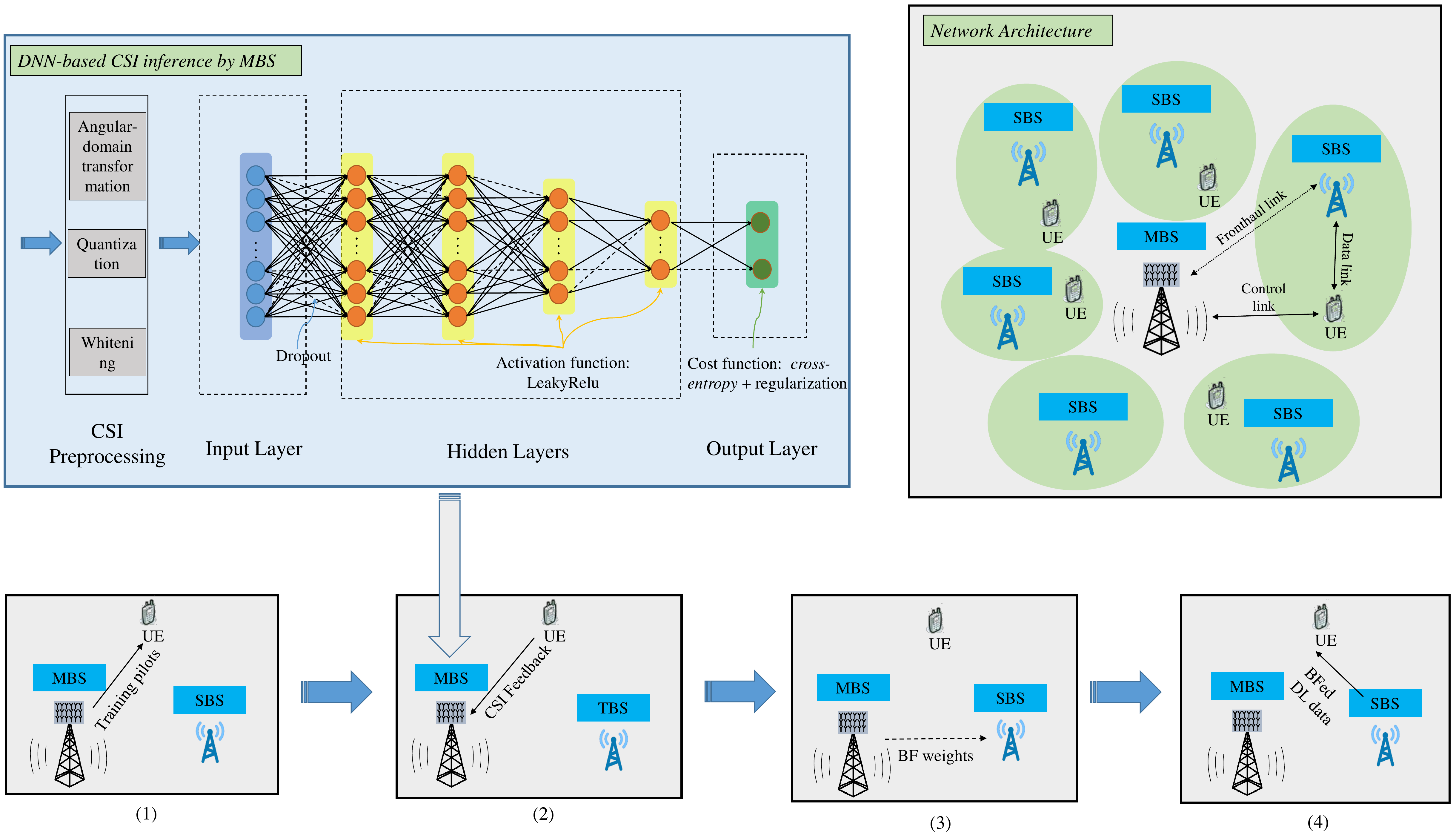}
	\caption{A UDN architecture with one MBS and several SBSs employing remote channel inference for beamforming where the DNN-based inference procedure is also shown at the bottom.}
	\label{fig_rbfi}
\end{figure*}

A simplified but concrete and insightful example is given here for illustration on the motivation for non-linear CSI structures. Let us assume that the wireless channels from a user to multiple BSs only have LoS MPCs (no angular spread), in which case the CSI can be perfectly characterized by a set of steering vectors with parameters of AoAs and amplitudes. Since AoAs are directly related to beamforming directions and contain most of the useful information of CSI in the LoS case, we consider the problem of inferring the AoA at an unknown BS given the AoAs at geographically separated BSs as an indication of (clearly) non-linear structures. In this formulation, the CSI exploitation is transformed to a parameter extraction problem and hence can be studied explicitly \cite{jiang18_globecom}. In particular, the CRLB of the CSI inference at a remote site given known CSI vectors at one or two MBSs, with this simplified model, scales inversely with $M^3$ with two known CSI vectors and $M$ with one known CSI, with $M$ being the number of antennas at an MBS; intriguingly the same scaling law is observed in a deep learning based implementation presented in \cite{jiang18_globecom}. We remark that in general scenarios, the complicated CSI structures due to multiple scattering, shadowing and so forth are often theoretically intractable and no explicit analysis is possible---sometimes even an accurate modeling is difficult, hence the model inadequacy.

\subsection{DNN-Based Remote CSI Inference}
The data-driven DNN approach has the advantage of model-free representation or function learning such that no explicit models of the complicated wireless channels are needed, at the expense of requiring large amounts of training data and corresponding computational power. Therefore, it is adopted to address the general CSI structure exploitation problem. As a first step towards this end, we identify two key ingredients in the success of applying DNNs to CSI data and then present case study results.

\emph{\textbf{CSI-Adapted DNN Feature Extraction:}} Feature engineering is key to conventional machine learning techniques, whereas the modern deep learning approach advocates end-to-end learning, eliminating the need for lengthy and application-dependent human-engineered feature extraction. Nonetheless, it is recognized that deep learning can also benefit from well-engineered features. In the case of CSI data as learning inputs and outputs, through extensive tests and analysis, we propose to leverage the rich knowledge in the wireless communication literature to enhance the feature extraction, more specifically 
adapting the CSI inputs to better incorporate with the DNN through input pre-processing. Particularly in our implementations, we have found the following feature engineering techniques useful and in fact improve the inference performance compared to end-to-end learning, such as transforming the CSI into the angular domain, which often exhibits sparsity, and whitening the inputs by taking the logarithms based on the reasoning that the DNN prefers Gaussian inputs and the CSI usually has log-normal large-scale fading coefficients.

\emph{\textbf{Statistical or Instantaneous CSI?}} The feasibility of non-linear CSI structures exploitation stems from what we refer to as \emph{CSI manifestation}. In essence, the structure of CSI manifests itself as the system dimension (spatial, frequency and time) grows, leading to the fact that the optimal representation learning problem, namely the mapping from CSI to physical wireless channels (minimum sufficient statistics), becomes better conditioned \cite{jiang18_globecom}. However, based on our experience, the current system dimension and computing power have not yet allowed full manifestation in the sense that one should not expect that the exact representation can be learned to predict instantaneous CSI; in other words, the function mapping learning of instantaneous CSI is infeasible. Instead, learning CSI statistics such as angular power spectrum are already very helpful in determining beamforming or user scheduling. Understanding this limitation is key for successful implementations since overshooting for instantaneous CSI inference in fact increases the variance of DNN training, resulting in slower training convergence to the global optimum.

\section{Case Studies}
\label{sec_cs}
We demonstrate the potential performance benefit of exploiting non-linear CSI structures by three case studies. The first one is a remote static CSI inference framework in ultra dense networks (UDNs), the second is time-series sequence-to-sequence CSI inference considering computing and communication delays, and the third is angular power spectrum inference for multi-user MIMO downlink user grouping. These cases present an intriguing observation which was previously deemed unlikely, that the CSI at a remote site with distance far larger than the carrier wavelength can be inferred by the CSI acquired at a local site.
\subsection[A]{Remote Static CSI Inference Based on DNN}
\begin{figure}[!t]
	\centering 
	{\includegraphics[width=0.5\textwidth]{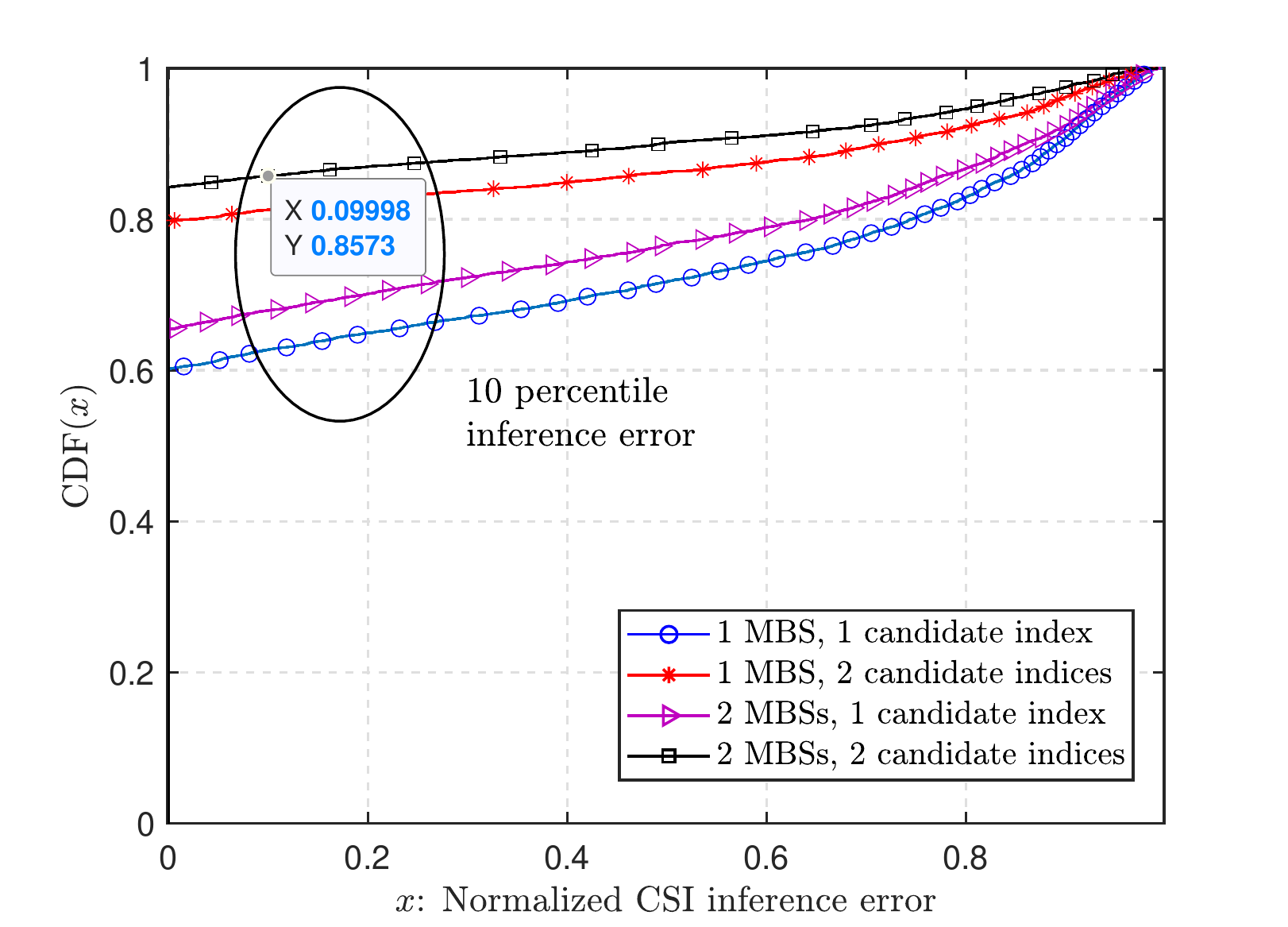}}
	\caption{Cumulative distribution function (CDF) of normalized inference error (normalized by the optimal quantized CSI in the codebook).}
	\label{fig_err}
\end{figure} 
A UDN is considered where there is one MBS and several SBSs as shown in Fig. \ref{fig_rbfi}. It is assumed that the number of antennas at the MBS is larger than that at the SBSs, such that the CSI estimation at the MBS contains more information and is thus leveraged to infer the CSIs at the SBSs. A high-level procedure of the remote beamforming inference framework is also shown in Fig. \ref{fig_rbfi}. In this framework where we focus on the FDD system, only the MBS sends downlink pilots and the CSIs at SBSs are inferred such that the pilot overhead in the network is significantly reduced. The users connect to the MBS and feed back the estimated downlink CSI. After that, one of the SBSs, which is decided by the MBS by selecting the best based on CSI estimation, will exploit beamforming technology to transmit downlink data to the user. Since the SBSs do not probe the channel and hence the CSI between the SBS and the user cannot be obtained directly, we resort to the data-driven deep neural network (DNN) based approach to make the channel inference \cite{chen17}. 

As discussed in Section \ref{sec_linear}, the main challenge faced in exploiting the non-linear structures in CSI is the difficulty in extracting an arbitrary dependency from CSIs. To address this issue, the DNNs are adopted since they are proven to be universal approximators capable of approximating any measureable function to any desired degree of accuracy, under the condition that the network parameters can be learned effectively. The training samples for the DNN are obtained by offline channel estimations for both the MBS and SBSs at sampling points using, for example, pilot-aided channel training, and updated to re-train the model when the inference error increases. The CSI at the MBS is input to the DNN, which is pre-processed by angular transformation and quantization. The DNN adopted in this work has three hidden layers and each has $100$ neurons. The CSI between the SBS and user is quantized based on a DFT codebook for convenience to adopt the supervised learning framework. The output of the DNN is the soft decision index in the codebook for the inferred CSI. The CSI is generated by a ray-tracing propagation software and the CSI inference based beamforming performance is shown in Fig. \ref{fig_err}. In order to increase the robustness of CSI inference, we consider adopting CSI at multiple MBSs as input to the DNN. The output of DNN is also generalized to two candidate indices and the decision among the two is left for the SBS. It is shown that the inferred CSI at approximately $85\%$ of overall test points has less than $10\%$ normalized inference error \cite{chen17}.

\subsection[B]{Time-Sequence CSI Learning with Neural Attentions}
\begin{figure}[!t]
	\centering
	{\includegraphics[width=0.48\textwidth]{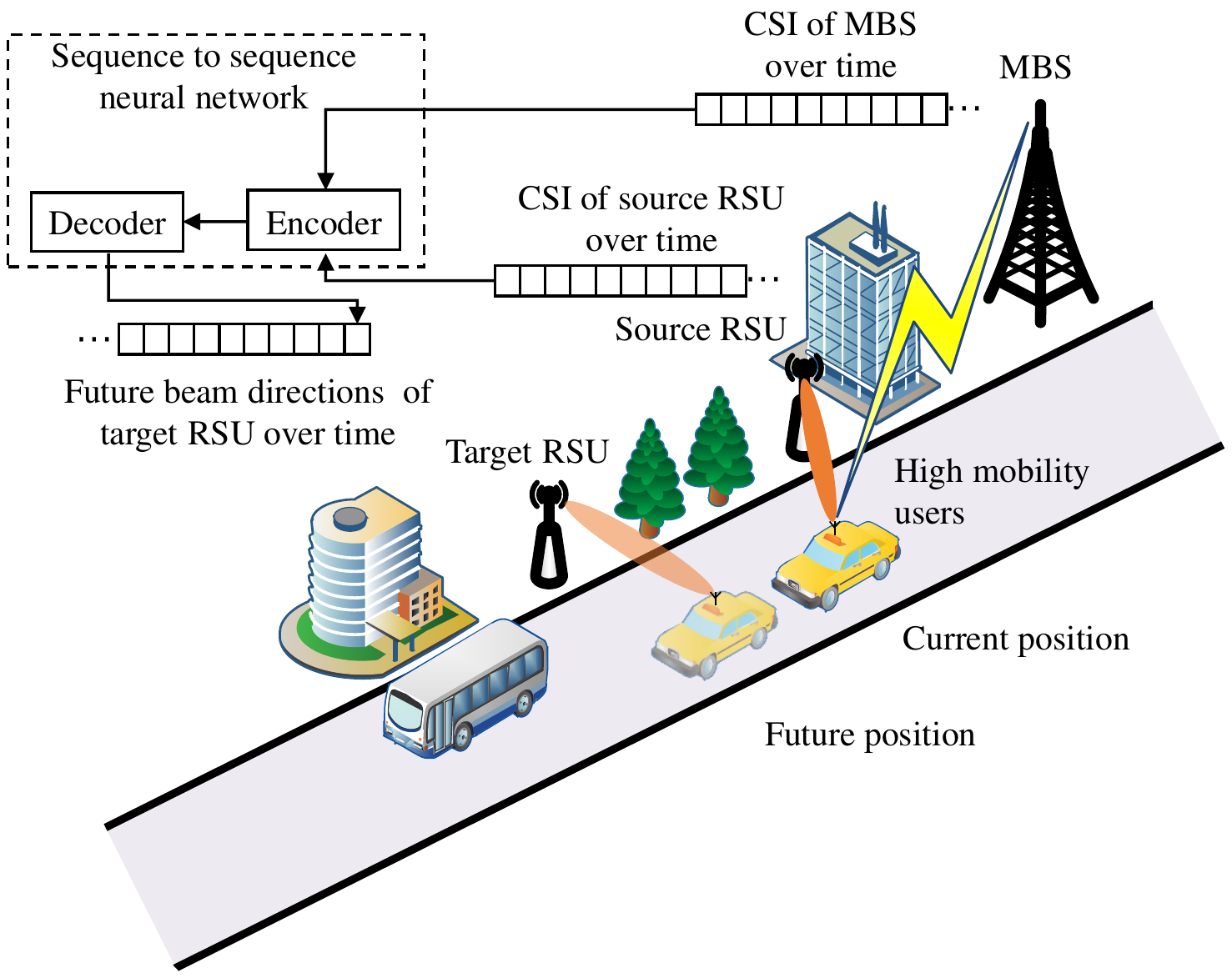}}	
	\caption{Schematic of the proposed time-sequence CSI inference framework.}
	\label{fig_rnnsce}
\end{figure}

The time-varying nature of CSI and possible inference delay, including computing, processing and backhaul transmission delays, bring up another challenge for aforementioned data-driven CSI learning framework, especially in a high mobility scenario. To cope with this issue, a modern neural network architecture, namely the recurrent neural network (RNN) is applied. One application scenario is shown in Fig. \ref{fig_rnnsce}, a vehicle will be handed over to the target road site unit (RSU) from the source RSU. In order to infer the CSI of the target RSU based on the known CSI from either the source RSU or the MBS, the inference delay has to be considered and properly compensated. In reality the delay is also varying due to backhaul queuing, and therefore a sequence-to-sequence learning model is adopted, with the input sequence being the CSI over some historical time slots at the source and the output at the target in some future time slots. The inference procedure is conducted in the source BS, and the inferred CSI is transmitted to the target RSU through the backhaul link. After receiving it, the target RSU measures the inference and transmission delay by using timestamps, and then establish the link to the vehicle based on the inferred CSI time sequence with the measured delay.

The performance of the proposed time-sequence CSI learning is evaluated on ray-tracing CSI data generated by Wireless Insite. It is found that \cite{chen18_globalsip} the RNN-based learning framework outperforms the location based (assuming LoS) beamforming (LO) method even with no positioning error and an approximately $10\%$ advantage over LO with a positioning error of $1$~m variance; compared with the optimal CSI from a DFT-based codebook, a $4.93\%$ performance loss is observed. Without time-sequence learning, the DNN architecture in Fig. \ref{fig_rbfi} can achieve an equivalent performance as an RNN-based learning framework only when the prediction delay is relatively small, but drops significantly when the delay grows, whereas an RNN-based framework can predict the future beam directions better.

\subsection[C]{Impacts on Downlink Multi-User MIMO Achievable Rates by APS Inference}
The aforementioned CSI inference can achieve significant beamforming gain for single-user scenarios, however, multi-user beamforming needs to consider the entire angular power spectrum (APS) information to counteract interference. One straightforward alternative solution is to allocate orthogonal resources for users, at the cost of losing spatial multiplexing gain. In this regard, the proposed CSI learning framework can be extended to infer the APSs of users at the SBS. Specifically, a DNN is trained, where the input and output are the APSs at MBS and SBS, respectively. Based on the inferred APS, users with small overlap in terms of APS can be grouped to transmit simultaneously since they are spatially separated; different groups use orthogonal resources. 

\begin{figure}[!t]
    \centering    
    {\includegraphics[width=0.5\textwidth]{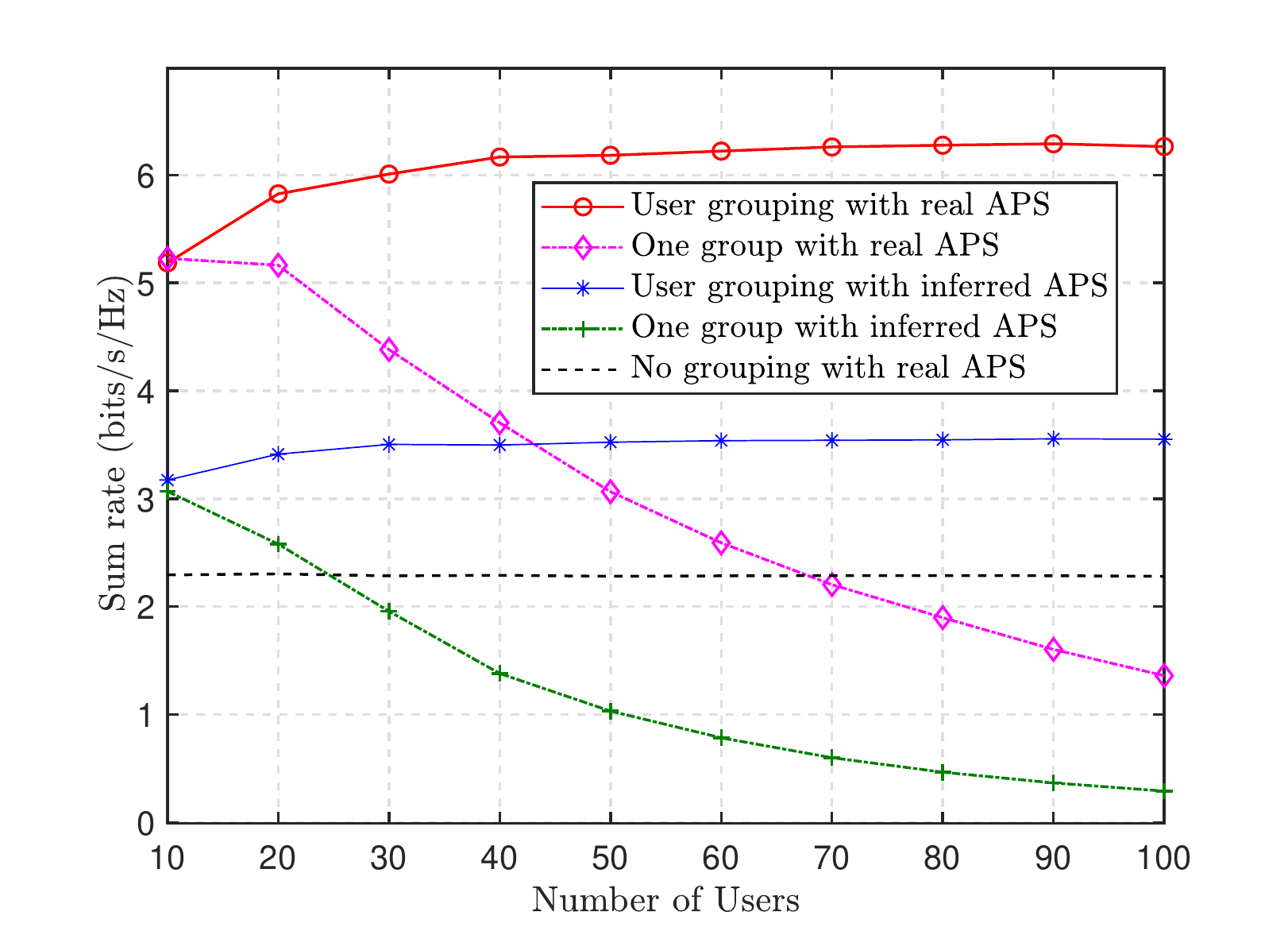}}
    \caption{Sum rate versus the number of users with SNR=$10$ dB.}
    \label{sim_pas}
\end{figure}

The performance of APS inference and user grouping is evaluated on ray tracing based CSI data. The SBS is equipped with a $1 \times 32$ antenna array, while the APS is a $1024$-dimensional vector for high angular resolution. A greedy graph coloring based user grouping with equal power allocation is adopted here to efficiently group users based on the inferred (or real) APSs. Two baseline algorithms are compared: the first is serving all users simultaneously in spite of the inference; the other is serving each user with orthogonal resources, based on DFT analog beamformer. The constrained achievable rate is evaluated, where the minimum signal-to-noise-ratio (SINR) threshold is $0.2$ (any SINR below the threshold is considered too low to achieve any throughput). As shown in Fig. \ref{sim_pas}, the proposed user grouping based on APS inference with much lower beam sweeping overhead can improve the sum rate compared with baseline algorithms, making it desirable for future mm-wave-based directional beamforming systems. 

\section{Concluding Remarks}
\label{sec_cl}
In this article, we systematically reviewed the state-of-the-art of CSI structure exploitation schemes and proposed to further investigate the non-linear structures. By theoretical feasibility and case studies based on realistic ray-tracing simulations, it was demonstrated that such structures exist and can be leveraged to further enhance the wireless communication system performance. Being model-free and data-driven, DNNs were recognized as necessary and powerful tools to deal with the inadequate channel modelling and untractable analysis in general CSI structure exploitation problems.

\emph{\textbf{Implications on System Design:}} The CSI structure has fundamental impacts on wireless system design. For instance, the emerging beamspace (beam-based, or angle-based) transmission is a direct consequence of the emergence of spatial linear CSI structure in M-MIMO implementation. It was shown in case studies that exploring non-linear CSI structures affects beamforming and pilot pattern designs. Moreover, the non-linear CSI structures also imply that cellular network operations can be enhanced, in the sense that a remote BS can infer the CSIs of sleeping BSs such that idle BSs can be put into sleeping mode to save energy without having to wake up to sniff the channels.

\emph{\textbf{Manifestations of CSI Structures:}} It is important to realize the difference between CSI structures and propagation channel structures. The wireless propagation channel is determined by the physical environment and does not depend on the system. On the other hand, CSI is a perception of the physical channel by the system and thus depends on system dimensions, e.g., antenna array size and bandwidth. As the system dimension increases and hence the granularity of CSI becomes finer and finer, the structure of perceived CSI begins to manifest itself; this trend will continue with higher and more diversified demands in future wireless systems, and the CSI structure will hence have increasingly prominent impacts on the system design.
\bibliographystyle{ieeetr}
\bibliography{csi}
\begin{IEEEbiographynophoto}{Zhiyuan Jiang}
	(S'12--M'15) received his B.E., Ph.D. degrees from the Electronic Engineering Department of Tsinghua University in 2010, 2015, respectively. He is currently an Associate Professor with the School of Communication and Information Engineering, Shanghai University. His main research interests include sequential decision making in wireless networks and massive MIMO systems.
\end{IEEEbiographynophoto}
\begin{IEEEbiographynophoto}{Sheng Chen}
	(S'16) received his Bachelor’s degree from Tsinghua University, Beijing, in 2016. Currently, he is studying at Tsinghua University for a Ph.D. degree. His research interests include channel estimation in massive MIMO systems, millimeter wave beamforming and machine learning applications in wireless communications.
\end{IEEEbiographynophoto}
\begin{IEEEbiographynophoto}{Andreas F. Molisch}
	(S'89--M'95--SM'00--F'05) is the Solomon Golomb -- Andrew and Erna Viterbi Chair Professor at the University of Southern California. He was previously at TU Vienna, AT\&T (Bell) Labs, Lund University, and Mitsubishi Electric Research Labs. His research interests are in wireless communications, with emphasis on propagation channels, multi–antenna systems, ultrawideband communications and localization, wireless video, and novel modulation formats. He is the author of 4 books, 19 book chapters, 230 journal papers, and numerous conference papers as well as 80 patents. He is Fellow of NAI, AAAS, IEEE, IET,  and Member of the Austrian Academy of Sciences, and recipient of numerous awards.  
\end{IEEEbiographynophoto}
\begin{IEEEbiographynophoto}{Rath Vannithamby}
	(S'90--M'00--SM'09) received his BS, MS, and PhD degrees in EE from the University of Toronto, Canada. He is a senior research scientist in Intel Labs, Intel Corporation, USA responsible for 5G+ research. He is a two times recipient of Intel Top Inventor award. Previously, he was a researcher at Ericsson. He is a Senior Member of IEEE. He was an IEEE Communications Society Distinguished Lecturer for 2014-2017. 
\end{IEEEbiographynophoto}
\begin{IEEEbiographynophoto}{Sheng Zhou}
	(S'06--M'12) received his B.S. and Ph.D. degrees in Electronic Engineering from Tsinghua University, China, in 2005 and 2011, respectively. He is currently an associate professor of Electronic Engineering Department, Tsinghua University. His research interests include cross-layer design for multiple antenna systems, vehicular networks, mobile edge computing, and green wireless communications.
\end{IEEEbiographynophoto}
\begin{IEEEbiographynophoto}{Zhisheng Niu}
	(M'98--SM'99--F'12) graduated from Beijing Jiaotong University, China, in 1985, and got his M.E. and D.E. degrees from Toyohashi University of Technology, Japan, in 1989 and 1992, respectively.  During 1992-94, he worked for Fujitsu Laboratories Ltd., Japan, and in 1994 joined with Tsinghua University, Beijing, China, where he is now a professor at the Department of Electronic Engineering. His major research interests include queueing theory, traffic engineering, mobile Internet, radio resource management of wireless networks, and green communication and networks.
\end{IEEEbiographynophoto}
\end{document}